\documentclass{article}
\usepackage{spconf,amsmath,graphicx}
\usepackage[backend=biber,style=ieee,doi=false,isbn=false]{biblatex}
\usepackage{amssymb}
\usepackage{amsthm}
\usepackage{cancel}
\usepackage{breqn}
\usepackage{multirow}
\usepackage{rotating}
\usepackage{verbatim}
\usepackage{textcomp,gensymb}
\usepackage[caption=false, font=footnotesize]{subfig}

\newcommand{\Expect}{{\rm I\kern-.5em E}}

\newcommand*\dif{\mathop{}\!\mathrm{d}}

\DeclareMathAlphabet{\mathitsf}{\encodingdefault}{\sfdefault}{m}{sl}
\DeclareMathAlphabet{\mathitbfsf}{\encodingdefault}{\sfdefault}{bx}{sl}
\usepackage{cool}
\newcounter{MYtempeqncnt}

\title{Channel Dependent Mutual Information in Index Modulations}
\name{Pol Henarejos$^{\star}$, Ana P\'erez-Neira$^{\star}$, Anxo Tato$^{\dagger}$, Carlos Mosquera$^{\dagger}$\thanks{This work is funded by projects MYRADA (TEC2016-75103-C2-2-R), ELISA (TEC2014-59255-C3-1-R) and TERESA (TEC2017-90093-C3-1-R).}}

\address{$^{\star}$ Centre Tecnol\`ogic de Telecomunicacions de Catalunya (CTTC), Castelldefels, Spain \\
	$^{\dagger}$ atlanTTic Research Center, University of Vigo, Spain\\
	Email: \{pol.henarejos,ana.perez\}@cttc.es$^{\star}$, \{anxotato,mosquera\}@gts.uvigo.es$^{\dagger}$}
\begin{document}
%
\maketitle
\begin{abstract}
Mutual Information is the metric that is used to perform link adaptation, which allows to achieve rates near capacity. The computation of adaptive transmission modes is achieved by employing the mapping between the Signal to Noise Ratio and the Mutual Information. Due to the high complexity of the computation of the Mutual Information, this process is performed off-line via Monte Carlo simulations, whose results are stored in look-up tables. However, in Index Modulations, such as Spatial Modulation or Polarized Modulation, this is not feasible since the constellation and the Mutual Information are channel dependent and it would require to compute this metric at each time instant if the channel is time varying. In this paper, we propose different approximations in order to obtain a simple closed-form expression that allows to compute the Mutual Information at each time instant and thus, making feasible the link adaptation.
\end{abstract}
\section{Introduction}
\label{sec:intro}

Link Adaptation in modern communications is performed by computing the Effective Signal to Noise (SNR) Mapping (ESM) based on Mutual Information (MI-ESM) \cite{Tao2011,Cheema2014,Hosseini2016}. For instance, the work described in \cite{emd2008} describes the procedure of computing MI-ESM in Single-Input Single-Output systems for IEEE 802.16e standard. Analogously, authors of \cite{Latif2012} describe the MI-ESM algorithm for Long Term Evolution (LTE) networks. All of these works have in common the computation of the Mutual Information (MI), which involves an expectation of a function of a Random Variable (RV) without closed-form solution.

In the literature, the computation of the expectation of MI is performed off-line via Monte Carlo simulations and the results are stored into a look-up table (LUT). After this step, the received SNR of each symbol within a codeblock or frame is mapped to the LUT to obtain the MI corresponding to the SNR. 

In Index Modulations (IM), such as Spatial Modulation \cite{DiRenzo2014} or Polarized Modulation \cite{Henarejos2015a}, the information is transmitted not only with a fixed constellation, such as Quadrature Amplitude Modulation (QAM), but also with the channel hops. Due to the dependence on the channel, the MI computation cannot be performed off-line since the expressions contain the channel realization \cite{Tato17}. The solution is to compute the MI curve in each time instant, depending on the channel realization. Due to the high computational complexity of MI computation, this approach is not feasible.

This paper presents closed-form expressions based on different order approximations of the MI of IM. Based on the works \cite{Yang2008,Rajashekar2014,Henarejos2017}, which compute the capacity of IM, we aim at solving the difficulty of finding a closed-form expression of MI. Thanks to this expression, we are able to compute the MI at each time instant with much less computational complexity and making the problem of adaptive IM affordable. Hence, the MI estimated is used to select the Modulation and Coding Scheme in the link adaptation algorithm process.

\section{System Model and Mutual Information}
Given a discrete time instant, the IM over an arbitrary Multiple-Input Multiple-Output (MIMO) channel realization, with $t$ inputs and $r$ outputs, is defined as
\begin{equation}
\mathbf{y}=\sqrt{\gamma}\mathbf{H}\mathbf{x}+\mathbf{w},
\label{eq:sysmod}
\end{equation}
where $\mathbf{y}\in\mathbb{C}^{r}$ is the received vector, $\gamma$ is the average SNR, $\mathbf{x}=\mathbf{l}s$, $\mathbf{l}$ is the all-zero vector except at position $l$ that is 1,  $\mathbf{H}=\left[\mathbf{h}_1\,\ldots\,\mathbf{h}_{t}\right]\in\mathbb{C}^{r\times t}$ is the channel matrix, $l\in\left[1,t\right]$ is the hopping index, $s\in\mathbb{C}$ is the complex symbol from the constellation $\mathcal{S}$. The AWGN noise is modeled as vector $\mathbf{w}\in\mathbb{C}^{r}\sim\mathcal{CN}\left(\mathbf{0},\mathbf{I}_r\right)$. In other words, $\mathbf{x}$ has only one component different from zero ($l$th component) and its value is $s$; that is, the transmitted symbol hops among the different channels. 

Differently from previous works, in this paper we do not analyze the statistics of $\mathbf{H}$, as we are only interested in the MI given a channel realization. $\mathbf{H}$ models the effects and specific impairments of the employed domain (spatial, polarization, frequency, etc.).
\begin{figure*}[!t]
	\normalsize
	\addtocounter{equation}{1}
	\begin{equation}
	\begin{split}
	h(s,l|\mathbf{y})&=-\sum_{s\in\mathcal{S}}\sum_{l=1}^t\int_{\mathcal{Y}}f_{\mathitsf{S},\mathitsf{L},\mathitbfsf{Y}}(s,l,\mathbf{y})\log_2\left(f_{\mathitsf{S},\mathitsf{L}|\mathitbfsf{Y}}(s,l,\mathbf{y})\right)\dif \mathbf{y}=\sum_{s\in\mathcal{S}}\sum_{l=1}^t\int_{\mathcal{Y}}f_{\mathitsf{S},\mathitsf{L},\mathitbfsf{Y}}(s,l,\mathbf{y})\log_2\left(\frac{f_{\mathitbfsf{Y}}(\mathbf{y})}{f_{\mathitsf{S},\mathitsf{L},\mathitbfsf{Y}}(s,l,\mathbf{y})}\right)\dif \mathbf{y}\\
	&=\sum_{s\in\mathcal{S}}\sum_{l=1}^t\int_{\mathcal{Y}}f_{\mathitbfsf{Y}|\mathitsf{S},\mathitsf{L}}(\mathbf{y},s,l)p_{\mathitsf{S}}(s)p_{\mathitsf{L}}(l)\times\log_2\left(\frac{\sum_{s'\in\mathcal{S}}\sum_{l'=1}^tf_{\mathitbfsf{Y}|\mathitsf{S},\mathitsf{L}}(\mathbf{y},s',l')p_{\mathitsf{S}}(s=s')p_{\mathitsf{L}}(l=l')}{f_{\mathitbfsf{Y}|\mathitsf{S},\mathitsf{L}}(\mathbf{y},s,l)p_{\mathitsf{S}}(s)p_{\mathitsf{L}}(l)}\right)\dif \mathbf{y}\\
	&=\frac{1}{tS}\sum_{s\in\mathcal{S}}\sum_{l=1}^t\Expect_{\mathitbfsf{Y}|\mathitsf{S},\mathitsf{L}}\left\{\log_2\left(\frac{\sum_{s'\in\mathcal{S}}\sum_{l'=1}^tf_{\mathitbfsf{Y}|\mathitsf{S},\mathitsf{L}}(\mathbf{y},s',l')}{f_{\mathitbfsf{Y}|\mathitsf{S},\mathitsf{L}}(\mathbf{y},s,l)}\right)\right\}
	\label{eq:condhdef}
	\end{split}
	\end{equation}
	\hrulefill
	\vspace*{4pt}
\end{figure*}
\addtocounter{equation}{-2}

Since the transmitted vector is determined by $(s,l)$, it is possible to rewrite \eqref{eq:sysmod} as
\begin{equation}
\mathbf{y}=\sqrt{\gamma}\mathbf{h}_ls+\mathbf{w}.
\label{eq:imsysmod}
\end{equation}
\addtocounter{equation}{1}
Thus, the MI between the received signal and $(s,l)$ is expressed as
\begin{equation}
\begin{split}
I(\mathbf{y};s,l)&=I(\mathbf{y};s|l)+I(\mathbf{y};l)\\
&=H(s|l)-h(s|l,\mathbf{y})+H(l)-h(l|\mathbf{y})\\
&=H(s)+H(l)-h(s,l|y)
\end{split}
\label{eq:defMI}
\end{equation}
where the third equality assumes that $s$ and $l$ are independent RV, $H(\mathitsf{X})=-\sum_{x\in\mathitsf{X}}p_{\mathitsf{X}}(x)\log_2\left(p_{\mathitsf{X}}(x)\right)$ is the entropy of $\mathitsf{X}$ and $h(\mathitsf{X})=-\int_{-\infty}^{\infty}f_{\mathitsf{X}}(x)\log_2\left(f_{\mathitsf{X}}(x)\right)\dif x$ is the differential entropy of $\mathitsf{X}$. Note that, in contrast to \cite{Henarejos2017}, where the capacity is obtained, in our case the symbol $s$ is not maximized and belongs to a particular constellation.

The entropy of $s$ and $l$ is expressed as $H(s)=\log_2S$ and $H(l)=\log_2t$, where $S$ is the number of symbols defined in the constellation. The expression of the differential entropy $h(s,l|\mathbf{y})$ is denoted in \eqref{eq:condhdef}, where $\mathcal{Y}$ is the domain of $\mathbf{y}$, $\Expect_{\mathitsf{X}}\left\{\cdot\right\}$ is the expectation of $\mathitsf{X}$, $f_{\mathitsf{S},\mathitsf{L},\mathitbfsf{Y}}(s,l,\mathbf{y})$ is the joint probability density function (pdf) of $s$, $l$ and $\mathbf{y}$, $f_{\mathitbfsf{Y}|\mathitsf{S},\mathitsf{L}}(\mathbf{y},s,l)$ is the conditional pdf of $\mathbf{y}$ conditioned to $s$ and $l$, $f_{\mathitbfsf{Y}}(\mathbf{y})$ is the pdf of $\mathbf{y}$, $p_{\mathitsf{S}}(s)=1/S$ and $p_{\mathitsf{L}}(l)=1/t$ are the probabilities of symbol $s$ and index $l$, respectively, $\int_{\mathcal{Y}}\dif\mathbf{y}\doteq\int_{\mathcal{Y}_1}\cdots\int_{\mathcal{Y}_r}\dif y_1\ldots\dif y_r$, and $\mathcal{Y}_i$ is the domain of the $i$th component of $\mathbf{y}$.

The pdf of $\mathbf{y}$ conditioned to $s$ and $l$ is obtained by assuming $s$ and $l$ to be deterministic in \eqref{eq:imsysmod}. In this case, it is clear that $\mathbf{y}$ is a multivariate complex Gaussian RV, with mean equal to $\sqrt{\gamma}\mathbf{h}_ls$ and identity covariance. Thus, the conditioned pdf is expressed as
\begin{equation}
f_{\mathitbfsf{Y}|\mathitsf{S},\mathitsf{L}}(\mathbf{y},s,l)=\frac{1}{\pi^r}e^{-\|\mathbf{y}-\sqrt{\gamma}\mathbf{h}_ls\|^2}.
\label{eq:fysl}
\end{equation}
Note that we assume that $s$ and $l$ are equiprobable. By substituting \eqref{eq:fysl} in \eqref{eq:condhdef}, the expectation can be described as
\begin{equation}
\begin{split}
&\Expect_{\mathitbfsf{Y}|\mathitsf{S},\mathitsf{L}}\left\{\log_2\left(\frac{\sum_{s'\in\mathcal{S}}\sum_{l'=1}^tf_{\mathitbfsf{Y}|\mathitsf{S}',\mathitsf{L}'}(\mathbf{y},s',l')}{f_{\mathitbfsf{Y}|\mathitsf{S},\mathitsf{L}}(s,l,\mathbf{y})}\right)\right\}\\
&=\Expect_{\mathitbfsf{\mathitbfsf{W}}}\left\{\log_2\left(\sum_{s'\in\mathcal{S}}\sum_{l'=1}^te^{-\gamma\left\|\mathbf{h}_ls+\frac{\mathbf{w}}{\sqrt{\gamma}}-\mathbf{h}_{l'}s'\right\|^2+\gamma\left\|\frac{\mathbf{w}}{\sqrt{\gamma}}\right\|^2}\right)\right\}\\
&=\Expect_{\mathitbfsf{\mathitbfsf{W}'}}\left\{\log_2\left(\sum_{s'\in\mathcal{S}}\sum_{l'=1}^te^{-\gamma\left(\left\|\mathbf{h}_ls-\mathbf{h}_{l'}s'+\mathbf{w'}\right\|^2-\left\|\mathbf{w'}\right\|^2\right)}\right)\right\},
\label{eq:exwp}
\end{split}
\end{equation}
where $\mathitbfsf{W}'\sim\mathcal{CN}\left(\mathbf{0},\frac{1}{\gamma}\mathbf{I}\right)$ and, thus, the conditioned RV $\mathitbfsf{Y}|\mathitsf{S},\mathitsf{L}\equiv\mathitbfsf{W}'$.

Computing \eqref{eq:exwp} is achieved numerically by generating a very large number of realizations of $\mathitbfsf{W'}$ and averaging the results via Monte Carlo simulations. However, this can only be feasible in scenarios where fixed constellations are employed. In the case of IM, the constellation depends on the channel realization. Hence, the expectation has to be calculated at each time instant, requiring high computational complexity and making the problem of link adaptation unaffordable. Our approach overcomes this problem, since it does not require off-line computations and presents closed-form expressions.

Once $f_{\mathitbfsf{W'}}$ is defined, we apply the same procedure as described in \cite{Henarejos2017}, which uses the Taylor Series Expansion (TSE) to approximate the expectation of a function by its moments. The central moments of $\mathitbfsf{W}'$ are defined by
\begin{equation}
\begin{split}
\mu_{\mathitsf{W}'_{i,\Re}}=\mu_{\mathitsf{W}'_{i,\Im}}&=0\\
\vartheta_{\mathitsf{W}'_{i,\Re}}^n=\vartheta_{\mathitsf{W}'_{i,\Im}}^n&=\begin{cases}(n-1)!!\frac{1}{(2\gamma)^{\frac{n}{2}}}= &\text{if \textit{n} is even}\\0&\text{if \textit{n} is odd}\end{cases},
\end{split}
\end{equation}
where $\mathitsf{W}'_{i,\Re}$ and $\mathitsf{W}'_{i,\Im}$ are the real and imaginary parts of the $i$th component of the RV $\mathitbfsf{W}'$. By assuming that

\begin{figure*}[!t]
	\normalsize
	\setcounter{MYtempeqncnt}{\value{equation}}
	\addtocounter{equation}{2}
	\begin{equation}
	\begin{split}
	I(\mathbf{y};s,l)&=\log_2(tS)-\frac{1}{tS}\sum_{s\in\mathcal{S}}\sum_{l=1}^t\log_2\left(\mathcal{D}_{sl}\right)-\frac{1}{tS}\sum_{s\in\mathcal{S}}\sum_{l=1}^t\sum_{n=1}^{\infty}\frac{1}{(2\gamma)^{n}(2n)!!}\sum_{m=1}^r\left(\frac{\partial^{2n}g_{sl}}{\partial w_{m,\Re}^{'2n}}(\boldsymbol{\mu}_{\mathitbfsf{W}'})+\frac{\partial^{2n}g_{sl}}{\partial w_{m,\Im}^{'2n}}(\boldsymbol{\mu}_{\mathitbfsf{W}'})\right)
	\label{eq:trueI2tay}
	\end{split}
	\end{equation}
	\hrulefill
	\vspace*{4pt}
\end{figure*}
\begin{figure*}[!t]
	\normalsize
	\setcounter{MYtempeqncnt}{\value{equation}}
	\begin{equation}
	\begin{split}
	&\sum_{m=1}^r\left(\frac{\partial^{2}g_{sl}}{\partial w_{m,\Re}^{'2}}(\boldsymbol{\mu}_{\mathitbfsf{W}'})+\frac{\partial^{2}g_{sl}}{\partial w_{m,\Im}^{'2}}(\boldsymbol{\mu}_{\mathitbfsf{W}'})\right)=\frac{4\gamma\sum_{s'\in\mathcal{S}}\sum_{l'=1}^t\mathcal{D}_{sl,s'l'}\log_2\left(\mathcal{D}_{sl,s'l'}^{-1}\right)}{\mathcal{D}_{sl}}\\
	&-\frac{(2\gamma)^2}{\log(2)}\sum_{m=1}^r\left[\left(\frac{\sum_{s'\in\mathcal{S}}\sum_{l'=1}^t\left(x_{m,s'l',\Re}-x_{m,sl,\Re}\right)\mathcal{D}_{sl,s'l'}}{\mathcal{D}_{sl}}\right)^2+\left(\frac{\sum_{s'\in\mathcal{S}}\sum_{l'=1}^t\left(x_{m,s'l',\Im}-x_{m,sl,\Im}\right)\mathcal{D}_{sl,s'l'}}{\mathcal{D}_{sl}}\right)^2\right]\\
	&=\frac{-4\gamma\log_2\left(\mathfrak{G}_{sl}\left(\mathcal{D}_{sl,s'l'}^{\mathcal{D}_{sl,s'l'}}\right)\right)}{\mathfrak{A}_{sl}\left(\mathcal{D}_{sl,s'l'}\right)}-\frac{(2\gamma)^2}{\log(2)\mathcal{D}_{sl}^2}\sum_{m=1}^r\left(\mathcal{D}_{m,sl,\Re}^2+\mathcal{D}_{m,sl,\Im}^2\right),
	\label{eq:trueapprox2}
	\end{split}
	\end{equation}
	\setcounter{MYtempeqncnt}{\value{equation}}
	\hrulefill
	\vspace*{4pt}
\end{figure*}
\begin{figure*}[!t]
	\normalsize
	\setcounter{MYtempeqncnt}{\value{equation}}
	\begin{equation}
	\begin{split}
	&I_{(2)}(\mathbf{y};s,l)\simeq\log_2\left(\frac{tS}{\mathfrak{G}\left(\mathcal{D}_{sl}\right)}\right)+\mathfrak{A}\left(\frac{\log_2\left(\mathfrak{G}_{sl}\left(\mathcal{D}_{sl,s'l'}^{\mathcal{D}_{sl,s'l'}}\right)\right)}{\mathfrak{A}_{sl}\left(\mathcal{D}_{sl,s'l'}\right)}+\frac{\gamma}{\log(2)\mathcal{D}_{sl}^2}\sum_{m=1}^r\left(\mathcal{D}_{m,sl,\Re}^2+\mathcal{D}_{m,sl,\Im}^2\right)\right)
	\label{eq:MI2}
	\end{split}
	\end{equation}
	\setcounter{MYtempeqncnt}{\value{equation}}
	\addtocounter{equation}{-5}
	\hrulefill
	\vspace*{4pt}
\end{figure*}
\begin{equation}
g_{sl}\left(\mathbf{w}'\right)=\log_2\left(\sum_{s'\in\mathcal{S}}\sum_{l'=1}^te^{-\gamma\left(\left\|\mathbf{x}_{sl}-\mathbf{x}_{s'l'}+\mathbf{w'}\right\|^2-\left\|\mathbf{w'}\right\|^2\right)}\right),
\label{eq:defG}
\end{equation}
we define the TSE of function $g_{sl}(\mathbf{w}')$ in the vicinity of $\boldsymbol{\mu}_{\mathitbfsf{W}'}$ as $g_{sl}(\mathbf{w}')=T(g_{sl},\mathbf{w}',\boldsymbol{\mu}_{\mathitbfsf{W}'})=P_N(g_{sl},\mathbf{w}',\boldsymbol{\mu}_{\mathitbfsf{W}'})+R_N(g_{sl},\mathbf{w}',\boldsymbol{\xi})$, where $P_N$ is the Taylor polynomial of degree $N$ and $R_N$ is the remainder term of degree $N$. Thus, the expectation of \eqref{eq:exwp} is equal to 
\begin{equation}
\begin{split}
&\Expect_{\mathitbfsf{W}'}\left\{T(g_{sl},\mathbf{w}',\boldsymbol{\mu}_{\mathitbfsf{W}'})\right\}=g_{sl}\left(\boldsymbol{\mu}_{\mathitbfsf{W}'}\right)\\
&+\sum_{n=1}^{\infty}\frac{1}{(2\gamma)^n(2n)!!}\sum_{m=1}^r\left(\frac{\partial^{2n}g_{sl}}{\partial w_{m,\Re}^{'2n}}(\boldsymbol{\mu}_{\mathitbfsf{W}'})+\frac{\partial^{2n}g_{sl}}{\partial w_{m,\Im}^{'2n}}(\boldsymbol{\mu}_{\mathitbfsf{W}'})\right)\\
&\doteq \mathcal{P}_N\left(g_{sl},\mathbf{w}',\boldsymbol{\mu}_{\mathitbfsf{W}'}\right)+\mathcal{R}_N\left(g_{sl},\mathbf{w}',\boldsymbol{\xi}\right),
\label{eq:truetayex}
\end{split}
\end{equation}
where $\boldsymbol{\xi}\in[\boldsymbol{\mu}_{\mathitbfsf{W}'},\mathbf{w}']$ and
\addtocounter{equation}{3}
\begin{equation}
\begin{split}
&\mathcal{P}_N\left(g_{sl},\mathbf{w}',\boldsymbol{\mu}_{\mathitbfsf{W}'}\right)=\Expect_{\mathitbfsf{W}'}\left\{P_N\left(g_{sl},\mathbf{w}',\boldsymbol{\mu}_{\mathitbfsf{W}'}\right)\right\}=g_{sl}\left(\boldsymbol{\mu}_{\mathitbfsf{W}'}\right)\\
&+\sum_{n=1}^{\lfloor N/2\rfloor}\frac{1}{(2\gamma)^n(2n)!!}\sum_{m=1}^r\left(\frac{\partial^{2n}g_{sl}}{\partial w_{m,\Re}^{'2n}}(\boldsymbol{\mu}_{\mathitbfsf{W}'})+\frac{\partial^{2n}g_{sl}}{\partial w_{m,\Im}^{'2n}}(\boldsymbol{\mu}_{\mathitbfsf{W}'})\right)\\
&\mathcal{R}_N\left(g_{sl},\mathbf{w}',\boldsymbol{\xi}\right)=\Expect_{\mathitbfsf{W}'}\left\{R_N\left(g_{sl},\mathbf{w}',\boldsymbol{\xi}\right)\right\}.
\end{split}
\end{equation}
Hereinafter, for the sake of clarity, we introduce the following definitions:
\begin{equation}
\begin{split}
\mathbf{x}_{sl}&\doteq\mathbf{h}_ls\\
\mathcal{D}_{sl,s'l'}&\doteq e^{-\gamma\|\mathbf{x}_{sl}-\mathbf{x}_{s'l'}\|^2}\\
\mathcal{D}_{sl}&\doteq \sum_{s'\in\mathcal{S}}\sum_{l'=1}^t\mathcal{D}_{sl,s'l'}=\sum_{s'\in\mathcal{S}}\sum_{l'=1}^te^{-\gamma\|\mathbf{x}_{sl}-\mathbf{x}_{s'l'}\|^2}.
\end{split}
\end{equation}

The first term of \eqref{eq:truetayex} is described as
\begin{equation}
\begin{split}
g_{sl}(\boldsymbol{\mu}_{\mathitbfsf{W}'})&=\log_2\left(\sum_{s'\in\mathcal{S}}\sum_{l'=1}^te^{-\|\mathbf{x}_{sl}-\mathbf{x}_{s'l'}\|^2}\right)\\
&=\log_2\left(\mathcal{D}_{sl}\right).
\label{eq:g0}
\end{split}
\end{equation}

Thus, by using \eqref{eq:truetayex}, \eqref{eq:g0} and substituting them into \eqref{eq:defMI}, then the MI can be expressed in a closed-form as in \eqref{eq:trueI2tay} and it can be approximated by considering additional terms. The simplest expression is the first order approximation, which is obtained by omitting the third term in \eqref{eq:trueI2tay}. Consequently, the first order approximation is denoted by
\begin{equation} 
\begin{split}
I_{(1)}(\mathbf{y};s,l)&\simeq\log_2(tS)-\frac{1}{tS}\sum_{s\in\mathcal{S}}\sum_{l=1}^t\log_2\left(\mathcal{D}_{sl}\right)\\
&=\log_2\left(\frac{tS}{\mathfrak{G}\left(\mathcal{D}_{sl}\right)}\right),
\label{eq:MI1}
\end{split}
\end{equation}
where $\mathfrak{G}\left(\mathcal{D}_{sl}\right)$ and $\mathfrak{A}\left(\mathcal{D}_{sl}\right)$ are the geometric and arithmetic mean, respectively, i.e., $\mathfrak{G}\left(\mathcal{D}_{sl}\right)=\left(\prod_{s\in\mathcal{S}}\prod_{l=1}^t\mathcal{D}_{sl}\right)^{\frac{1}{tS}}$ and $\mathfrak{A}\left(\mathcal{D}_{sl}\right)=\frac{1}{tS}\sum_{s\in\mathcal{S}}\sum_{l=1}^t\mathcal{D}_{sl}$.

The second order approximation involves the second derivative of $g_{sl}$ at $\mathbf{x}_{sl}$. Thus, after some mathematical manipulations, the second term is expressed as \eqref{eq:trueapprox2}, where
\begin{equation}
\begin{split}
\mathcal{D}_{m,sl,\Re}&=\sum_{s'\in\mathcal{S}}\sum_{l'=1}^t\left(x_{m,s'l',\Re}-x_{m,sl,\Re}\right)\mathcal{D}_{sl,s'l'}\\
\mathcal{D}_{m,sl,\Im}&=\sum_{s'\in\mathcal{S}}\sum_{l'=1}^t\left(x_{m,s'l',\Im}-x_{m,sl,\Im}\right)\mathcal{D}_{sl,s'l'}
\end{split}
\end{equation}
and
\begin{equation}
\begin{split}
\mathfrak{A}_{sl}\left(\mathcal{D}_{sl,s'l'}\right)&=\frac{1}{tS}\sum_{s'\in\mathcal{S}}\sum_{l'=1}^t\mathcal{D}_{sl,s'l'}\\
\mathfrak{G}_{sl}\left(\mathcal{D}_{sl,s'l'}^{\mathcal{D}_{sl,s'l'}}\right)&=\left(\prod_{s'\in\mathcal{S}}\prod_{l'=1}^t\mathcal{D}_{sl,s'l'}^{\mathcal{D}_{sl,s'l'}}\right)^{\frac{1}{tS}}
\end{split}
\end{equation}
are the arithmetic and geometric means over $s'$ and $l'$ by keeping $s$ and $l$ fixed. Hence, by plugging \eqref{eq:trueapprox2} in \eqref{eq:trueI2tay}, the second order approximation of MI is described by \eqref{eq:MI2}.

\subsection{Bounds of approximated Mutual Information}
\label{sect:bounds}
TSE applied to the expectation of a function of a RV allows to express it as a function of its moments instead of the RV; thus, making more efficient the computation by successive approximations. An important remark is that the expectation of TSE is lower or upper bounded by the first order approximation, depending on its convexity or concavity, respectively. 

In our case, this can be proven by examining the convexity of \eqref{eq:defG} and applying the Jensen's inequality, which results that the expectation of TSE is lower bounded by \eqref{eq:g0}. 

This can be proven by using the Jensen's inequality as follows
\begin{equation}
\begin{split}
\mathcal{P}_1\left(f,\mathbf{x},\boldsymbol{\mu}_{\mathitbfsf{X}}\right)=f\left(\boldsymbol{\mu}_{\mathitbfsf{X}}\right)&=f\left(\Expect_{\mathitsf{X}}\left\{\boldsymbol{x}\right\}\right)\leq\Expect_{\mathitbfsf{X}}\left\{f(\mathbf{x})\right\}.
\end{split}
\end{equation}
Note that, due to the minus sign in \eqref{eq:defMI}, the lower bound of Jensen's inequality becomes an upper bound, which is increased by the factor $\log_2(tS)$ and averaged by $t$ and $S$.

\section{Results}
In this section, we illustrate the results derived from the previous sections. We compare the performance of first and second order approximations, i.e., \eqref{eq:MI1} and \eqref{eq:MI2}, respectively, by simulating the curves of MI with the integral-based expression \eqref{eq:condhdef}, \eqref{eq:defMI}.
\begin{figure}[!ht]
	\centering
	\subfloat[][$2\times 2$, $S=4$]{\includegraphics[width=0.45\linewidth,clip=true]{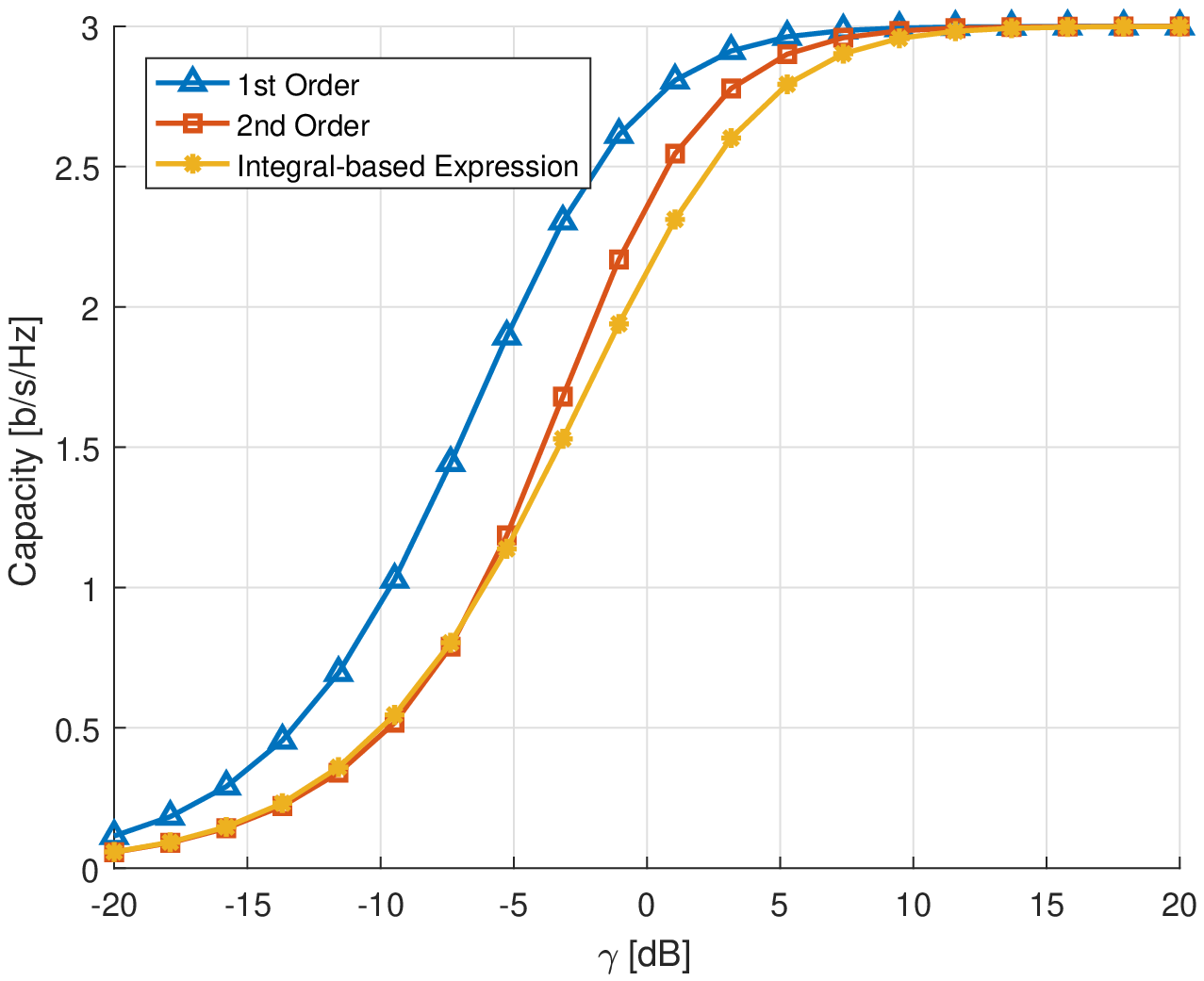}\label{fig:th2x2_rx}}\hfill
	\subfloat[][$2\times 2$, $S=16$]{\includegraphics[width=0.45\linewidth,clip=true]{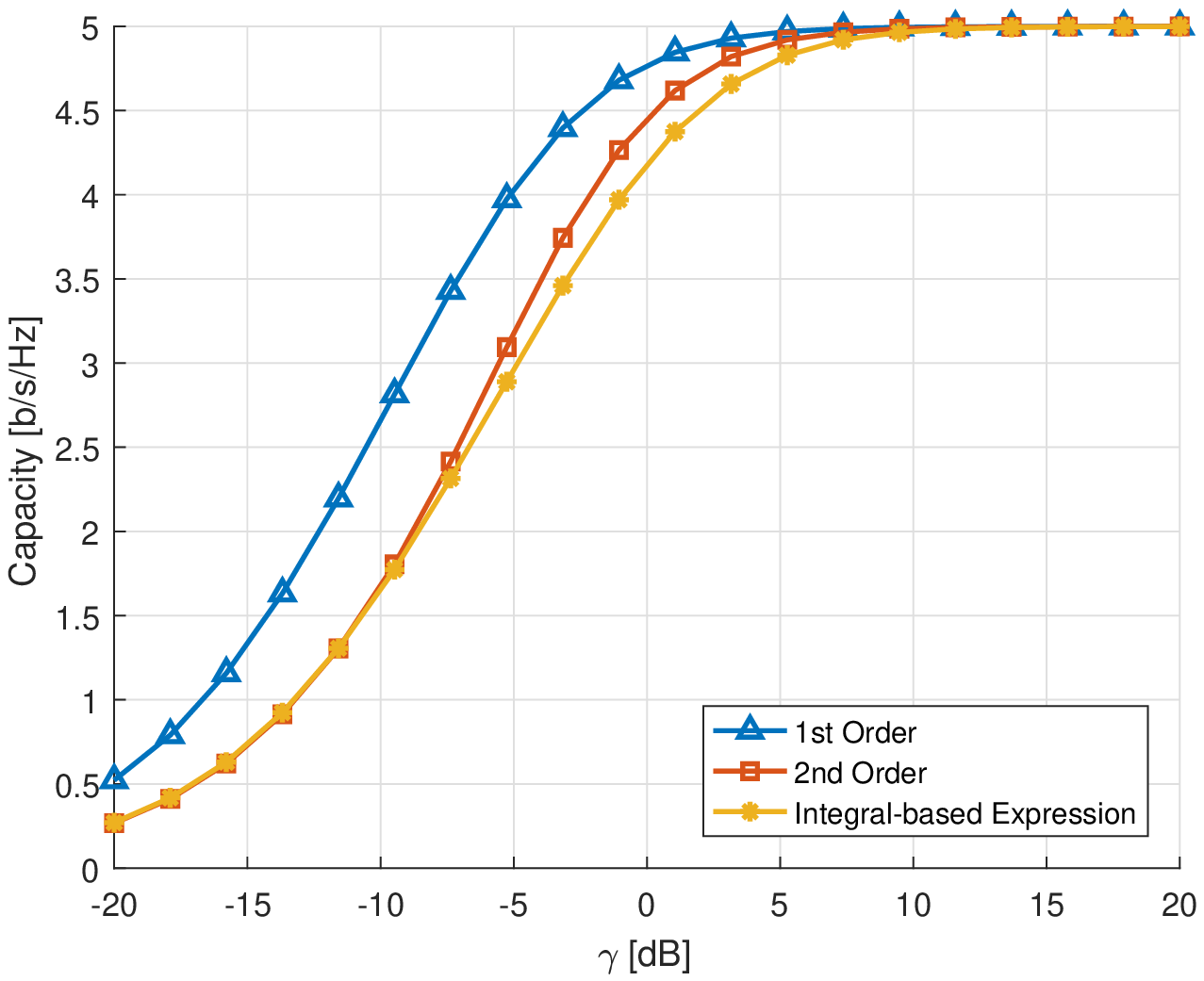}\label{fig:th2x4_rx}}\hfill
	\subfloat[][$4\times 4$, $S=4$]{\includegraphics[width=0.45\linewidth,clip=true]{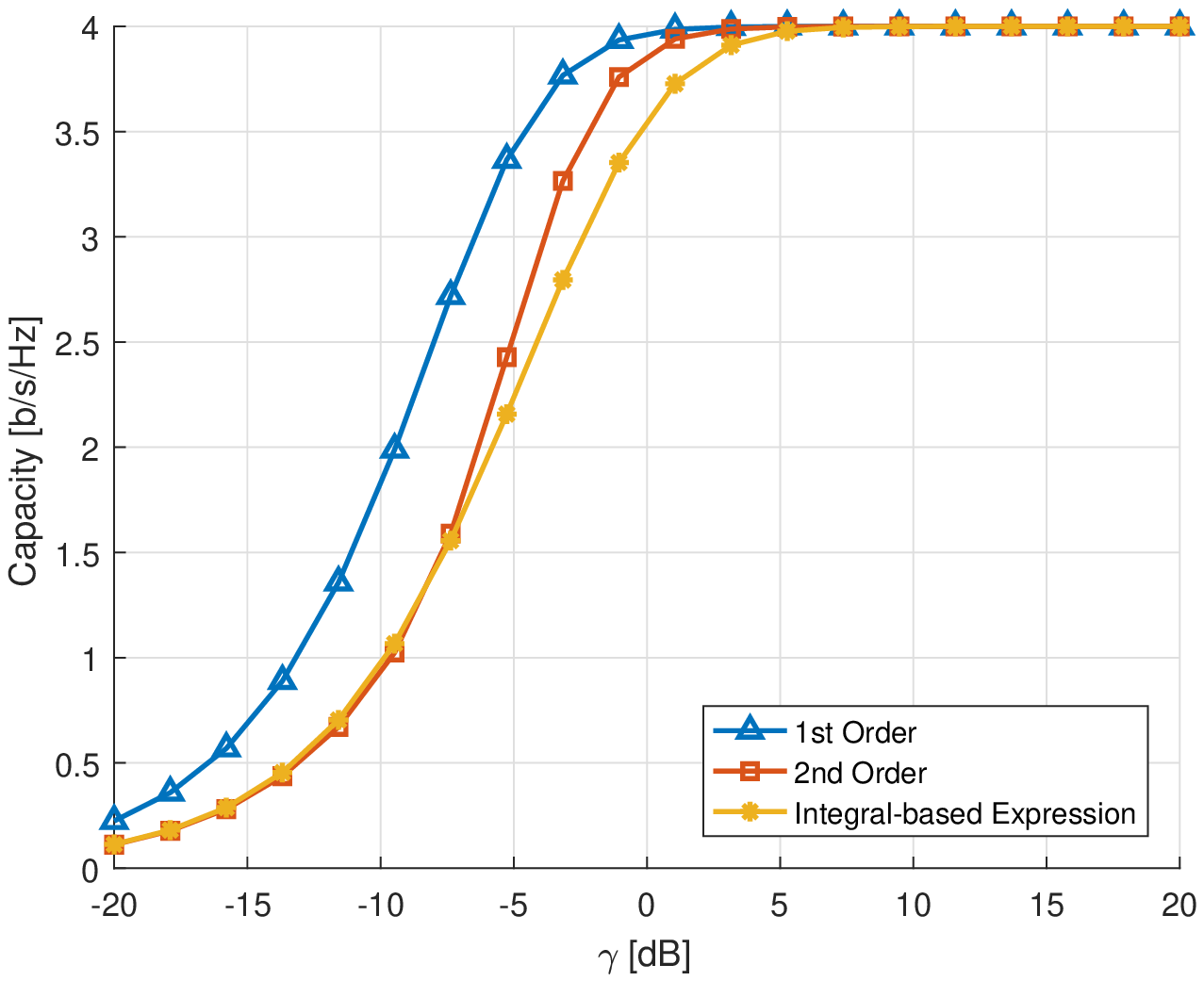}\label{fig:th4x4_rx}}\hfill
	\subfloat[][$4\times 4$, $S=16$]{\includegraphics[width=0.45\linewidth,clip=true]{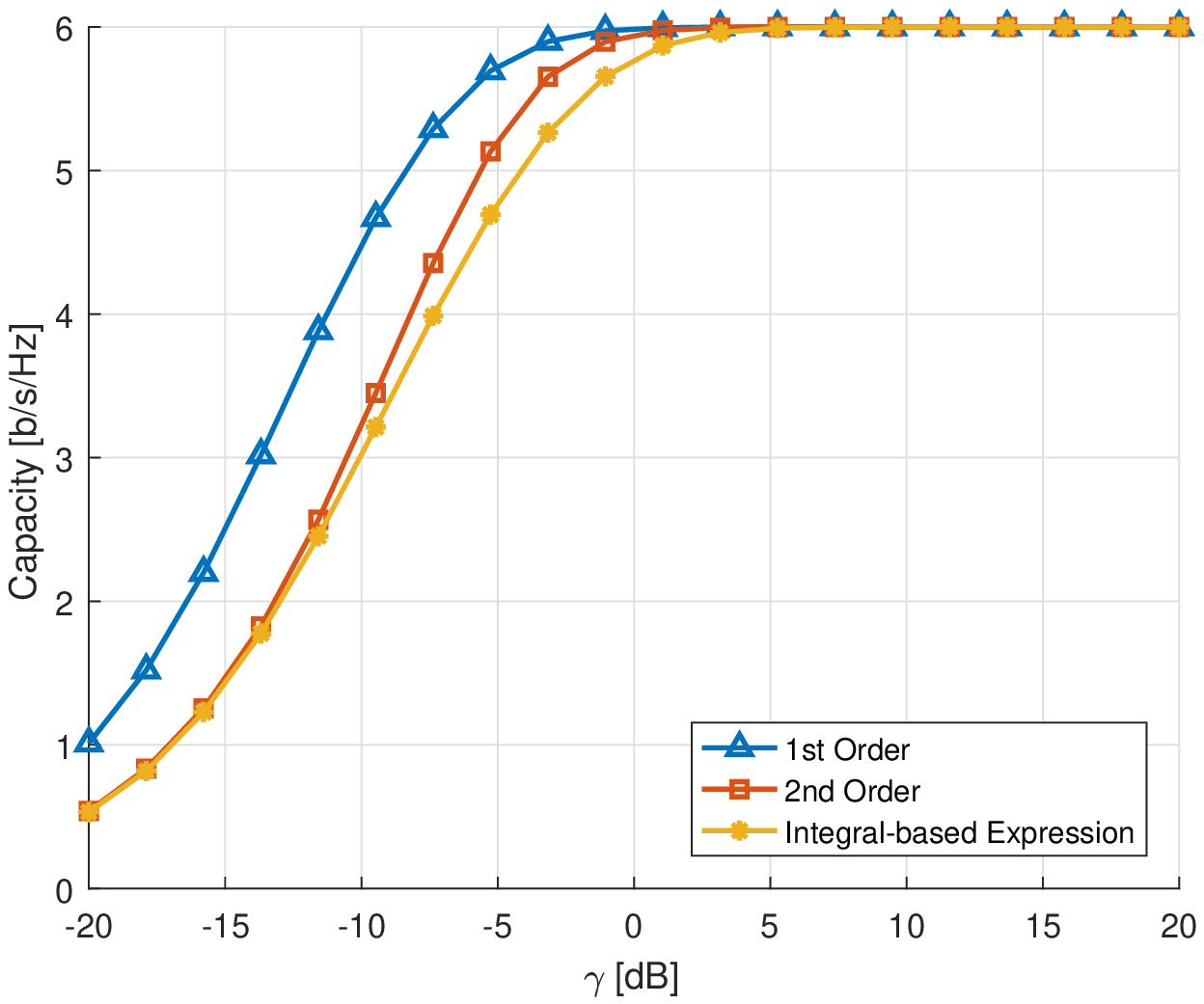}\label{fig:th4x8_rx}}\hfill
	\caption{Comparison of the MI for different order approximations and the integral-based expression, i.e., \eqref{eq:MI1}, \eqref{eq:MI2} and \eqref{eq:defMI}, respectively.}
	\label{fig:mis}
\end{figure}

In this simulation, we generate $10^3$ independent channel realizations following a Rayleigh distribution and average the results to obtain a single smooth curve. Note that we do not average over noise realizations since we obtained mathematical expressions that are not functions of a noise RV. We also depict different input/outputs configurations and different constellations. Particularly, we consider QPSK and 16-QAM constellations.

Fig. \ref{fig:mis} illustrates the MI of first and second order approximations, \eqref{eq:MI1} and \eqref{eq:MI2}, respectively, compared with the integral-based expression, \eqref{eq:condhdef}, \eqref{eq:defMI}. First, as we denoted in Section \ref{sect:bounds}, the first order approximation is, at the same time, the upper bound of the integral-based expression. Additionally, we can observe that, as expected, the second order approximation produces tighter curve compared with the first order approximation.

\section{Conclusions}
In this paper we introduce the problem of implementing link adaptation in Index Modulations, such as Spatial Modulation or Polarized Modulation, where the information is modulated with fixed constellations and dynamic channel hops. If the channel is time varying, it is unaffordable to compute the Mutual Information at each time instant. With our approach it is possible to obtain a smooth curve by using closed-form expressions, decreasing the computational complexity and allowing to perform the link adaptation. Finally, we depict the first and second order approximations compared with integral-based expression for several configurations and constellation size.

\vfill\pagebreak

\section{REFERENCES}
\label{sec:refs}

\printbibliography[heading=none]

\end{document}